\documentclass[reprint,
nofootinbib,
 amsmath,amssymb,
 aps,
]{revtex4-2}

\usepackage[utf8]{inputenc} % allow utf-8 input
\usepackage[T1]{fontenc}    % use 8-bit T1 fonts
\usepackage{hyperref}       % hyperlinks
\usepackage{url}            % simple URL typesetting
\usepackage{booktabs}       % professional-quality tables
\usepackage{amsfonts}       % blackboard math symbols
\usepackage{nicefrac}       % compact symbols for 1/2, etc.
\usepackage{microtype}      % microtypography
\usepackage{graphicx}
\usepackage{xcolor, etoolbox}
\usepackage{amsmath,amsfonts,amsthm} % Math packages
\usepackage{mathtools}
\usepackage{ragged2e}
\usepackage{braket}      % Dirac Bra-Ket notation
\usepackage[english]{babel} % English language/hyphenation
\graphicspath{{./figures/}} % Set graphics path
\usepackage{enumitem}
\usepackage{xspace}
\usepackage{dsfont}
\usepackage{bm}% bold math
\usepackage{bbm}% bold math
\usepackage{dcolumn}% Align table columns on decimal point

\usepackage[colorinlistoftodos]{todonotes}
\usepackage{mathrsfs}

\newcommand{\Tr}{\text{Tr}}

\renewcommand{\vec}[1]{\boldsymbol{#1}}

\hypersetup{
    colorlinks,
    linkcolor=blue,
    citecolor=blue,
    urlcolor=blue,
    breaklinks=true 
}

\DeclarePairedDelimiter\abs{\lvert}{\rvert}

\theoremstyle{remark}

\begin{document}

\title{Wehrl entropy, entropic uncertainty relations and entanglement}

\author{Stefan Floerchinger}
 \email{stefan.floerchinger@thphys.uni-heidelberg.de}
\author{Tobias Haas}
 \email{t.haas@thphys.uni-heidelberg.de}
\author{Henrik M\"{u}ller-Groeling}
 \email{h.mueller-groeling@thphys.uni-heidelberg.de}
 \affiliation{Institut f\"{u}r Theoretische Physik, Universit\"{a}t Heidelberg, \\ Philosophenweg 16, 69120 Heidelberg, Germany}

\begin{abstract}
The Wehrl entropy is an entropy associated to the Husimi quasi-probability distribution. We discuss how it can be used to formulate entropic uncertainty relations and for a quantification of entanglement in continuous variables. We show that the Wehrl-Lieb inequality is closer to equality than the usual Białynicki-Birula and Mycielski entropic uncertainty relation almost everywhere. Furthermore, we show how a Wehrl mutual information can be used to obtain a measurable perfect witness for pure state bipartite entanglement, which additionally provides a lower bound on the entanglement entropy.
\end{abstract}

\maketitle

%%%%%%%% Document %%%%%%%%

\section{Introduction}
\label{sec:Introduction}
Concepts of quantum information theory play an increasingly important role for describing and understanding complex quantum systems. Starting point is usually the von Neumann or quantum entropy associated to a state or density matrix $\rho$ \cite{vonNeumann1955,Nielsen2010,Wilde2013,Bengtsson2017},
\begin{equation}
    S (\rho) = - \Tr \left\{ \rho \ln \rho\right\}.
    \label{eq:defvNEnt}
\end{equation}
It quantifies missing information or uncertainty encoded in the state $\rho$ \cite{Shannon1948,Jaynes1963}. Despite its importance in quantum information theory, the von Neumann entropy $S(\rho)$ has a major drawback: it can typically not be accessed directly in experiments. A computation of the von Neumann entropy requires detailed knowledge of the quantum state $\rho$, which is oftentimes out of reach. While there are experimental approaches to reconstruct the full state $\rho$ through quantum state tomography \cite{Nielsen2010}, the effort for these techniques scales unfavorably with system size. Alternatively, quantum entropies can be measured in (relatively small) quantum systems with finite dimensional Hilbert space $\text{dim} \mathcal{H} < \infty$ by considering multiple copies of the quantum system and using quantum many-body interference \cite{Abanin2012,Islam2015}.

For infinite dimensional quantum states as they arise for example in continuous variable systems or quantum field theory it is substantially more difficult to access or constrain quantum entropies experimentally.

In this work, we investigate a somewhat different approach and consider the Wehrl entropy $S_{\text{W}} (\rho)$, which is an entropy of the Husimi $Q$-distribution, as a quasi-classical measure for missing information in quantum mechanical phase space \cite{Wehrl1978,Wehrl1979}. While the density operator $\rho$ and the Husimi $Q$-distribution contain both the full information about the state of the quantum system (see e.\ g.\ ref.\ \cite{Schleich2001}), the latter has the advantage that it can be accessed experimentally (see e.\ g.\ \cite{Kunkel2019} for recent developments in spinor Bose-Einstein condensates). Therefore, we suggest a more regular use of the Wehrl entropy $S_\text{W} (\rho)$ when analyzing tabletop experiments from an information theoretic perspective.

In this work, we focus on two major branches of modern quantum information theory, where the Wehrl entropy reveals its strength.

First, we discuss the role of Wehrl entropy in the context of entropic uncertainty relations (EURs). The first entropic uncertainty relation was introduced by Białynicki-Birula and Mycielski \cite{Bialynicki-Birula1975} based on previous works by Everett \cite{Everett1957}, Hirschmann \cite{Hirschman1957} and Beckner \cite{Beckner1975}. It is an uncertainty relation for a single pair of conjugate variables, namely position $x$ and momentum $p$, which may be distributed according to the probability density functions $f(x)=\braket{x | \rho | x}$ and $g(p)=\braket{p | \rho | p}$. This corresponds to the so-called homodyne detection protocol, where both variables are recorded separately by measuring in their corresponding eigenbases. Then, the Białynicki-Birula and Mycielski (BBM) entropic uncertainty relation reads \cite{Bialynicki-Birula2011,Coles2017,Hertz2019}
\begin{equation}
    h (f) + h (g) \geq \ln e \pi,
    \label{eq:BirulaEUR}
\end{equation}
where $h (f)$ is the classical differential entropy associated to the distribution $f(x)$ \cite{Cover2006}
\begin{equation}
    h (f) = - \int \text{d} x \, f(x) \ln f(x).
\end{equation}
It should be noted that the bound on the right-hand side of the inequality \eqref{eq:BirulaEUR} is state \textit{independent} whereas the left-hand side increases for mixed states due to concavity. Thus, one can expect the bound to be rather loose for mixed states. Nevertheless, the relation \eqref{eq:BirulaEUR} turns out to be stronger than and implies the well-known formulation of an uncertainty relation in terms of variances \cite{Robertson1929,Kennard1927,Heisenberg1927,Schroedinger1930}, which is discussed in detail for example in ref.\ \cite{Hertz2019}.

Another entropic uncertainty relation for continuous variables was found by Frank and Lieb (FL) \cite{Frank2012} based on previous work by Rumin \cite{Rumin2011}
\begin{equation}
    h (f) + h (g) \geq \ln 2 \pi + S(\rho).
    \label{eq:FrankEUR}
\end{equation}
Therein the von Neumann entropy $S(\rho)$ accounts for the \textit{mixedness} of the system's state $\rho$. Obviously $\ln 2 \pi$ alone is a less tight bound than the one in \eqref{eq:BirulaEUR}, but for highly mixed states one expects the bound in \eqref{eq:FrankEUR} to be closer to equality.

Also for the Wehrl entropy there exists an entropic uncertainty relation, which is known as the Wehrl-Lieb inequality (WL) \cite{Lieb1978,Carlen1991,Lieb2005,Grabowski1984,Ohya1993}
\begin{equation}
    S_{\text{W}} (\rho) \ge 1,
    \label{eq:WehrlLiebInequality1}
\end{equation}
where equality only holds for pure coherent state projectors $\rho_\alpha = \ket{\alpha} \bra{\alpha}$. 

It was shown that the latter relation does also imply the variance based uncertainty relation by Heisenberg and an entropic uncertainty relation of the form \eqref{eq:BirulaEUR} based on differential entropies of smeared-out distributions (the marginals of the Husimi $Q$-distribution, which are not the true marginals $f$ and $g$) \cite{Grabowski1984}.

One main concern of this work will be to investigate in which regimes the Wehrl-Lieb entropic uncertainty relation \eqref{eq:WehrlLiebInequality1} is tighter (closer to equality) than the relations in terms of differential entropies \eqref{eq:BirulaEUR} and \eqref{eq:FrankEUR}; this will be addressed in \autoref{sec:Monopartite}.

The second question we study is how the Wehrl entropy $S_{\text{W}} (\rho)$ can be used to make statements about entanglement in continuous variable systems. In general, detecting bipartite entanglement for a state $\rho$ which lives in an infinite dimensional Hilbert space $\mathcal{H}$ can be quite challenging \cite{Horodecki2009}. Thus, many criteria are concerned with analyzing the second-order moments \cite{Simon2000,Giovannetti2003,Guehne2004,Hyllus2006}, which is a powerful approach if the state $\rho$ is of Gaussian form (see e.\ g.\ refs. \cite{Weedbrook2012,Serafini2017,Berges2018} for a general discussion of Gaussian states). Incorporating also higher-order moments allows to define a sufficient condition for bipartite entanglement, but can be unmanageable in practice \cite{Schukin2005}. 

Another approach is to witness entanglement by violating conditions based on differential entropies \cite{Walborn2009} or differential conditional entropies \cite{Walborn2011,Frank2013,Furrer2014}, which are closely related to entropic uncertainty relations (see also ref.\ \cite{Huang2013,Bergh2021a,Bergh2021b}). They turn out to be successful in detecting many classes of entangled states, especially in the pure state case. 

What these approaches have in common is that they rely on a homodyne detection protocol. In \autoref{sec:Bipartite} we will present an approach based on a Wehrl mutual information\footnote{An approach solely based on Wehrl entropy is discussed in ref.\ \cite{Mintert2004}.}, which presumes a heterodyne detection scheme and turns out to witness every pure entangled state and is there furthermore able to bound the entanglement entropy from below. Moreover we discuss the role of the Wehrl conditional entropy, which differs from its fully quantum analog.

\textit{This paper is organized as follows}. We begin introducing coherent states and the Wehrl entropy of an $N$ mode system in \autoref{sec:Basics}. Then, we discuss the relations between the three entropic uncertainty relations \eqref{eq:WehrlLiebInequality1}, \eqref{eq:BirulaEUR} and \eqref{eq:FrankEUR} in \autoref{sec:Monopartite}. In particular, we consider common classes of states, such as number eigenstates or thermal states. This analysis allows us to show that the Wehrl-Lieb inequality is stronger than the relation by Białynicki-Birula and Mycielski almost everywhere in the examined regimes. In \autoref{sec:Bipartite} we consider a bipartite system of $N+M$ modes and construct a Wehrl mutual information from a Wehrl relative entropy, which witnesses every pure entangled bipartite state. The concept is applied to some well-known examples like the two-mode squeezed state or the class of N00N-states. Additionally, we discuss the Wehrl conditional entropy. Finally, we summarize our results and provide an outlook in \autoref{sec:Conclusion}.

\textit{Notation}. In this paper we work with natural units, i.e. $\hbar=k_\text{B}=1$ and we drop operator hats. Rather, we employ capital letters for observables $X$ and small letters for their eigenvalues $x$ and eigenvectors $\ket{x}$. Also, we use the symbol $S$ for the von Neumann entropy, $h$ for classical differential entropies and $S_{\text W}$ for Wehrl entropies with the underlying probability distribution $f(x)$ (or state $\rho$) as argument $h = h(f)$ (or $S = S(\rho)$). We denote conditional entropy by $S(A|B)$ and mutual information by $I(A:B)$.

\section{Coherent states and Wehrl entropy}
\label{sec:Basics}
We start from a set of $N$ continuous quantum variables $X_j$ and conjugate momenta $P_j$ and the (bosonic) commutation relation 
\begin{equation}
    [X_j, P_k] = i \delta_{jk}.
\label{eq:commuationRelationXP}
\end{equation}
While we use a notation as for positions of particles, the variables $X_j$ could also denote other quantities such as amplitudes of field modes. Note that we make no detailed assumptions about the Lagrangian or Hamiltonian at this point. 

We also introduce the annihilation operators
\begin{equation}
    a_j = \frac{1}{\sqrt{2}} (X_j + i P_j),
\end{equation}
and creation operators $a_j^\dagger = (a_j)^\dagger$ such that eq.\ \eqref{eq:commuationRelationXP} reads $[a_j, a_k^\dagger] = \delta_{jk}$. Coherent fields are defined as usual as eigenstates of the annihilation operators (see e.\ g.\ \cite{Schleich2001,Mandel2013}),
\begin{equation}
    a_j \ket{\vec{\alpha}} = \alpha_j \ket{\vec\alpha},
\end{equation}
with eigenvalues $\alpha_j \in \mathbb{C}$. They can be expressed in terms of number eigenstates (with ground state $| 0 \rangle$)
\begin{equation}
\ket{\vec n} = \prod_{j=1}^N \left\{ \frac{1}{\sqrt{n_j!}}(a_j^\dagger)^{n_j}  \right\} |0 \rangle,
\end{equation}
through 
\begin{equation}
    \ket{\vec \alpha} = \prod_{j=1}^N \left\{ \sum_{n_j=0}^{\infty} e^{- \frac{1}{2} \abs{\alpha_j}^2} \frac{\alpha_j^{n_j}}{\sqrt{n_j!}} \right\} \ket{\vec n}.
\end{equation}

For our purposes, it will often be convenient to parameterize the complex vector of eigenvalues $\vec{\alpha}$ in terms of position $\vec{x}$ and momentum $\vec{p}$,
\begin{equation}
    \vec{\alpha} = \frac{1}{\sqrt{2}} \left(\vec{x} + i \vec{p} \right).
    \label{eq:PhaseToPositionMomentum}
\end{equation}
Coherent states have many interesting properties. They are not orthogonal,
\begin{equation}
    \abs{ \braket{\vec{\alpha} | \vec{\beta}} }^2 = e^{- \abs{\vec{\alpha} - \vec{\beta}}^2},
\end{equation}
and span an (overcomplete) basis,
\begin{equation}
    \mathds{1} = \int \frac{\text{d}^{2N} \alpha}{\pi^N} \ket{\vec{\alpha}} \bra{\vec{\alpha}}.
    \label{eq:CoherentStatesIdentity}
\end{equation}
Moreover, they minimize the uncertainty principle, which suggests that they can be regarded as a favorable measurement basis for simultaneous measurements of $x$ and $p$. Note that there are other states minimizing the uncertainty principle (squeezed states). Coherent states are the only such states symmetric in the quadratures. In fact, we can construct a positive operator-valued measure (POVM), see e.\ g.\ ref.\ \cite{Holevo2011}, based on pure coherent state projectors
\begin{equation}
    E_{\vec{\alpha}} = \ket{\vec{\alpha}} \bra{\vec{\alpha}},
\end{equation}
which resolve the identity according to \eqref{eq:CoherentStatesIdentity}. The probability density for the outcome $\vec\alpha$ is given by
the Husimi $Q$-distribution \cite{Husimi1940,Schleich2001,Mandel2013},
\begin{equation}
    Q^{\rho} (\vec{\alpha}) = \Tr \left\{ \rho \, E_{\vec{\alpha}} \right\} = \braket{\vec{\alpha} | \rho | \vec{\alpha}},
\end{equation}
such that the probability to find $\vec \alpha$ in some interval is given by $Q^{\rho} (\vec{\alpha}) d^{2N}\alpha/\pi^N$. 

Based on the Husimi $Q$-distribution one can formally define a state after measurement, without recording the outcome,
\begin{equation}
    \rho_{\vec{\alpha}} = \int \frac{\text{d}^{2N} \alpha}{\pi^N} Q^{\rho} (\vec{\alpha}) \ket{\vec{\alpha}} \bra{\vec{\alpha}},
\end{equation}
and it is of Glauber $P$-form. 

The Husimi $Q$-distribution is a \textit{quasi}-probability distribution\footnote{For general literature on quasi-probability distributions see e.g. refs. \cite{Mandel2013,Zhang1990,Wigner1931,Glauber1963,Sudarshan1963,Lee1995}.} due to the non-orthogonality of coherent states, which causes the Husimi $Q$-distribution to violate Kolmogorov's third axiom, namely $\sigma$-additivity. In other words, two values $Q^{\rho} (\vec \alpha)$ and $Q^{\rho} (\vec \alpha')$ for $\vec \alpha \neq \vec \alpha'$ do not represent probabilities for mutually exclusive events. Nevertheless, $Q^{\rho} (\vec \alpha)$ is normalized to unity in the sense
\begin{equation}
    \int \frac{\text{d}^{2N} \alpha}{\pi^N} Q^{\rho} (\vec \alpha) = \Tr \{ \rho\} = 1. 
    \label{eq:HusimiQNormalization}
\end{equation}
What we have described here is known as the heterodyne detection protocol \cite{Schleich2001}. Intuitively, it corresponds to jointly measuring position $x$ and momentum $p$ with the highest precision allowed by Heisenberg's uncertainty relation. 

In contrast to the Wigner $W$-distribution, the Husimi $Q$-distribution is non-negative and bounded \cite{Cartwright1976,Lee1995}
\begin{equation}
    0 \le Q^{\rho} (\vec \alpha) \le 1,
\end{equation}
which allows to associate a (differential) entropy to it. This is known as the Wehrl entropy \cite{Wehrl1978,Wehrl1979}
\begin{equation}
\begin{split}
    S_\text{W} (\rho) & = - \int \frac{\text{d}^{2N} \alpha}{\pi^N} Q^{\rho} (\vec \alpha) \ln Q^{\rho} (\vec \alpha) \\
    & = - \int \frac{\text{d}^N x\, \text{d}^N p}{(2\pi)^N} Q^\rho(\vec x, \vec p) \ln Q^{\rho} (\vec x, \vec p).
\end{split}
\label{eq:WehrlEntropy}
\end{equation}
In the second equation we used the parametrization introduced in eq.\ \eqref{eq:PhaseToPositionMomentum}. 

The Wehrl entropy can be seen as a quasi-classical differential entropy in quantum mechanical phase space. It is bounded from below by the true von Neumann entropy (as expected for a coarse-grained notion of an entropy), but also bounded from above by the von Neumann entropy of the measured state\footnote{The second statement can be proven using the well-known Berezin-Lieb inequality \cite{Berezin1972}, see e.g. ref.\ \cite{Wehrl1979}.}
\begin{equation}
    S(\rho) \le S_{\text{W}} (\rho) \le S(\rho_{\vec \alpha}).
\end{equation}
Furthermore, it is uniquely minimized by any pure coherent state projector $\rho = \ket{\vec \alpha} \bra{\vec \alpha}$, in which case its value is $S_{\text{W}} = N$. This result is known as the Wehrl-Lieb inequality \cite{Lieb1978,Carlen1991,Lieb2014},
\begin{equation}
    S_{\text{W}} (\rho) \ge S_\text{W} \left(\ket{\vec \alpha}\bra{\vec \alpha} \right) = N.    
    \label{eq:WehrlLiebInequality}
\end{equation}
Rather recently, improvements of this bound were discussed in refs.\ \cite{dePalma2018a,dePalma2018b}.

Another property of the Wehrl entropy $S_{\text{W}} (\rho)$ is important when it comes to entanglement. In contrast to the von Neumann entropy $S(\rho)$, it is monotonous under partial trace, i.e. if we consider a bipartite quantum system $\mathcal{H} = \mathcal{H}_A \otimes \mathcal{H}_B$ with the reduced density operator $\rho_A = \Tr_B \{ \rho \}$, we have
\begin{equation}
    S_{\text{W}} (\rho_A) \le S_{\text{W}} (\rho).
    \label{eq:WehrlMonotonicity}
\end{equation}
This highlights the classical nature of Wehrl entropy. The analogous statement is oftentimes not true for the von Neumann entropy $S(\rho)$ due to entanglement \cite{Wehrl1978,Wehrl1979}. This already indicates that capturing entanglement with a Wehrl entropy may require a different line of reasoning compared to a von Neumann entropy.

\section{Monopartite quantum system}
\label{sec:Monopartite}
At first we consider a monopartite quantum system. In the following, we want to compare the differential entropies of the two phase space distributions $Q^\rho(x,p)/(2\pi)$ and $f(x)g(p)$ associated with the heterodyne and homodyne measurement protocol respectively. Let us first state the following relations
\begin{equation}
    \begin{split}
        h\left(\frac{Q^\rho}{2\pi}\right) &= -\int \text d x\, \text d p \frac{Q^\rho(x,p)}{2\pi} \ln \left(\frac{Q^\rho(x,p)}{2\pi}\right) \\
        &= S_{\text W}(\rho) + \ln (2\pi), \\
        h(fg) &= \int \text d x\, \text d p f(x) g(p) \ln(f(x)g(p)) \\
        &= h(f) + h(g).
    \end{split}
\end{equation}
The two differential entropies are related up to additive constants with $S_{\text W}(\rho)$ and $h(f) + h(g)$, for which we know entropic uncertainty relations.

Henceforth, our goal is to compare the uncertainty deficit, i.e. the distance to the bound of an entropic uncertainty relation, of these relations as an estimator for the strength of the statement. We refer to a lower uncertainty deficit as the \textit{tightness} of an uncertainty relation. It is clear that being close to equality is desirable for inequalities, however it should be noted here that this does not immediately lead to a formal implication between different uncertainty relations. In fact, the Wehrl-Lieb inequality is tight on a different set of states than the inequality by Białynicki-Birula and Mycielski.

To properly compare them, we rearrange the three inequalities such that their right-hand sides coincide. Thus, we have the following three relations
\begin{equation}
    \begin{rcases}
        \text{WL} &S_{\text{W}}(\rho) + \ln \pi \\
        \text{BBM} &h(f) + h(g) \\
        \text{FL} &h(f) + h(g) - S(\rho) + \ln e/2
    \end{rcases}
    \ge \ln e \pi.
    \label{eq:EURSummary}
\end{equation}
We begin our analysis with pure number eigenstates, then we investigate a mixture of the lowest two number eigenstates and finally consider thermal states. This analysis will allow us to state a conjecture about their relations for a general state $\rho$. If possible, results will be given analytically. Furthermore, the Husimi $Q$-distribution will be written down in terms of position $x$ and momentum $p$, i.e. using the parametrization \eqref{eq:PhaseToPositionMomentum}, in order to make the different (quasi)-probability distributions comparable. 

\subsection{Number eigenstates}
\label{sec:NumberEigenstates}
The density matrix of a pure number eigenstate
\begin{equation}
    \rho_n = \ket{n} \bra{n},
\end{equation}
has vanishing quantum entropy, $S(\rho_n) = 0$. As a consequence, the Frank-Lieb bound will be less tight than the Białynicki-Birula and Mycielski bound.

Furthermore, the probability density functions $f(x)$ and $g(p)$ are of equal shape,
\begin{equation}
    \begin{split}
        f_n (x) &= \abs{\psi_n (x)}^2 = \frac{1}{\sqrt{\pi} \, 2^n n!} \, H_n^2 (x) \, e^{-x^2},
        \label{eq:ProbabilitesFockStates}
    \end{split}
\end{equation}
where $H_n (x)$ are the commonly-known physicist's Hermite polynomials. The Husimi $Q$-distribution is given by 
\begin{equation}
    \begin{split}
     Q_n(x,p) &= \abs{\braket{n | \alpha}}^2 \\
     &= \frac{1}{2^n n!} \, \left( x^2 + p^2\right)^n \, e^{-\frac{1}{2}(x^2+p^2)}.
    \end{split}
    \label{eq:HusimiQFockStates}
\end{equation}
From these expressions all appearing entropies can be calculated in the following.

The differential entropies $h(f)$ and $h(g)$ appearing in the Białynicki-Birula and Mycielski as well as the Frank-Lieb uncertainty relation were investigated in refs.\ \cite{Dehesa1994,Majernik1996,Ruiz1997,Toranzo2019}. The main challenge in this computation is due to the Hermite polynomials $H_n (x)$. In principle the differential entropy is given by
\begin{equation}
    h(f_n) = \ln \left(\sqrt{\pi} \, 2^n n! \right) + n + \frac{1}{2} + \frac{1}{\sqrt{\pi} \, 2^n n!} E (H_n),
\end{equation}
where $E (H_n)$ is the so-called entropy of Hermite polynomial
\begin{equation}
    E (H_n) = -\int_{- \infty}^{+ \infty} d x \, H_n^2(x) \ln \left( H_n^2 (x) \right) e^{-x^2}.
\end{equation}
After a tedious and lengthy calculation one finds \cite{Ruiz1997,Toranzo2019}
\begin{equation}
    \begin{split}
     h(f_n) =& \ln \left(\sqrt{\pi} \, 2^n n! \right) + n + \frac{1}{2} + n \gamma \\
     & - 2 \sum_{i=1}^n x_{n,i}^2 \, {}_2 F_2 \left(1,1; \frac{3}{2}, 2; -x_{n,i}^2 \right) \\
     & + \sum_{k=1}^n \binom{n}{k} \frac{(-1)^k 2^k}{k} \sum_{i=1}^n {}_1 F_1 \left(k, \frac{1}{2}, - x_{n,i}^2 \right).
     \end{split}
\end{equation}
Therein ${}_2 F_2 (a,b;c,d;z)$ denotes a generalized hypergeometric function, ${}_1 F_1 (a,b;z)$ is Kummer's confluent hypergeometric function for $z \in \mathbb{C}$ and $x_{n,i}$ are the roots of the $n$-th Hermite polynomial $H_n (x^2)$. Also, we have introduced the Euler-Mascheroni constant $\gamma \approx 0.577$.

The latter expression can be evaluated numerically for all $n \in \mathbb{N}$. Furthermore, for $n \gg 1$ one can give a simple approximate formula for the differential entropy \cite{Ruiz1997}
\begin{equation}
    h (f_n) \approx \frac{1}{2} \left(-2 + \ln 2 \pi^2 n \right).
    \label{eq:FockDifferentialEntropyApprox}
\end{equation}
In contrast, the Wehrl entropy can be computed analytically due to the simple form of the Husimi $Q$-distribution for number eigenstates \eqref{eq:HusimiQFockStates}. Although this is a rather technical issue, one might appreciate the fact that computing a Wehrl entropy is much easier than computing a differential entropy in many cases.

\begin{figure*}[t!]
    \centering
    \includegraphics[height=0.23\textwidth]{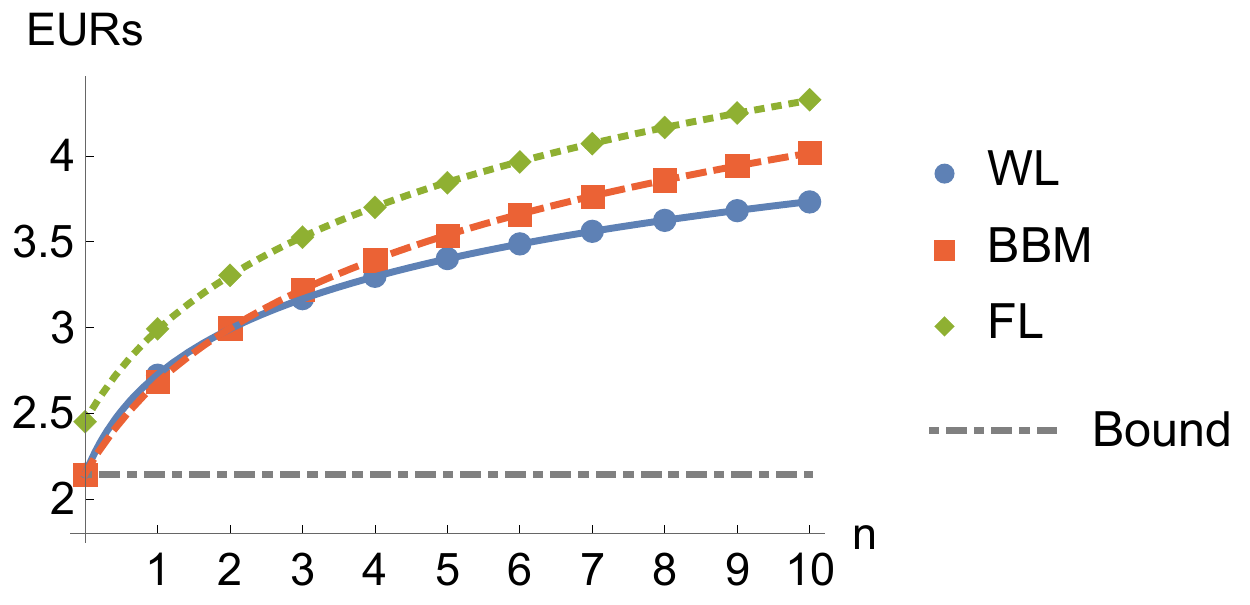}
    \includegraphics[height=0.22\textwidth]{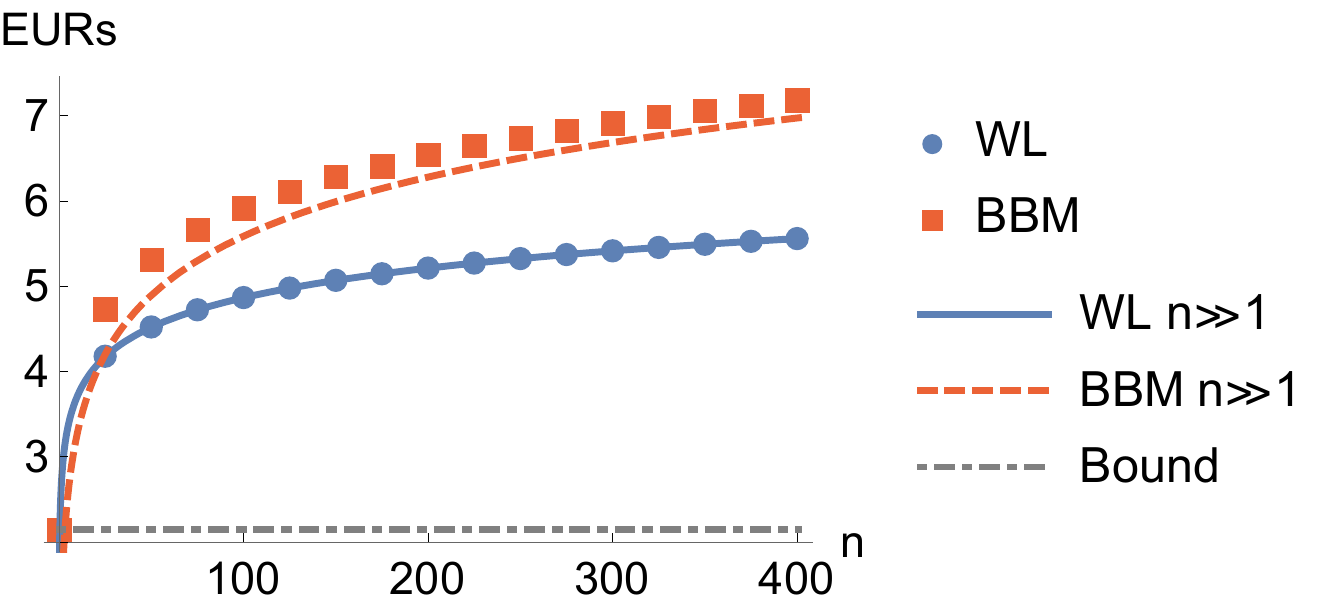}
    \caption{Left panel: Values of the three entropic uncertainty relations (EURs, cf. \eqref{eq:EURSummary}) as a function of the energy level $n$. One can see that the Białynicki-Birula and Mycielski (BBM) as well as the Wehrl-Lieb (WL) inequality are equivalent and tight for the ground state $n=0$. For $n=1$ we have $S_{\text{W}} (\rho_1) + \ln \pi \approx 2.72 > h (f_1) + h(g_1) \approx 2.69$, whereas for $n=2$ we get $S_{\text{W}} (\rho_2) + \ln \pi \approx 2.992 < h (f_2) + h(g_2) \approx 2.997$. Thus, the Wehrl-Lieb uncertainty relation is the strongest for $n \in \mathbb{N} \setminus \{1 \}$, i.e. almost everywhere. For pure states the bound of the Frank-Lieb (FL) relation becomes state-independent. In that case it is less tight than the relation by Białynicki-Birula and Mycielski. Right panel: Behavior of the Wehrl-Lieb and the Białynicki-Birula and Mycielski uncertainty relations for large $n$ (the latter was investigated numerically in \cite{Majernik1996}). The result for the Wehrl entropy is already valid for small $n$, whereas the result for the differential entropies is only valid for $n \gg 1$. This is again due to the fact that the Wehrl entropy is of such simple form.} 
    \label{fig:QHOEURPureStates}
\end{figure*}

Starting from \eqref{eq:HusimiQFockStates}, one can compute the Wehrl entropy for number eigenstates. After a straight forward exercise in Gaussian integration one finds (see also ref.\ \cite{Orlowski1999})
\begin{equation}
    S_{\text{W}} (\rho_n) = \ln{n!} + n + 1 + n\gamma - n \eta_n,
    \label{eq:FockWehrlEntr2}
\end{equation}
where $\eta_n$ is the $n$-th harmonic number 
\begin{equation}
    \eta_n = \sum_{k=1}^{n} \frac{1}{k},
\end{equation}
with $\eta_0 = 0$.

The result can be further simplified when considering large $n \gg 1$ and using the Stirling approximation formula 
\begin{equation}
    n! \approx \sqrt{2\pi n}\left(\frac{n}{e}\right)^n,
\end{equation}
which leads to
\begin{equation}
    S_{\text{W}}(\rho_n) \approx \frac{1}{2} \left( 1 + \ln{2\pi n} \right), \label{eq:FockWehrlApprox}
\end{equation}
for $n \gg 1$.

For the sake of a visual comparison we present all three bounds as a function of $n$ in the left panel of \autoref{fig:QHOEURPureStates}. The main result is that the Wehrl-Lieb relation is tightest for $n>1$. Moreover, we obtain the expected result for $n=0$, i.e. the Wehrl-Lieb as well as the Białynicki-Birula and Mycielski relations are tight and equivalent. In general, the Frank-Lieb relation is worst since we consider pure states.

Furthermore, it is interesting to compare the Wehrl-Lieb relation with the Białynicki-Birula and Mycielski relation for large $n \gg 1$ (cf. right panel of \autoref{fig:QHOEURPureStates}). From the asymptotic expressions \eqref{eq:FockDifferentialEntropyApprox} and \eqref{eq:FockWehrlApprox} we can read off the scaling behaviors for large $n \gg 1$, which gives $h(f_n) + h(g_n) \propto \ln n$ and $S_{\text{W}} (\rho_n) \propto 1/2 \, \ln n$. Therefore, the Wehrl-Lieb relation grows much slower than the Białynicki-Birula and Mycielski relation. Also, we observe that the Stirling approximation \eqref{eq:FockWehrlApprox} is close to the true result \eqref{eq:FockWehrlEntr2} also for small $n \sim 1$, while $\eqref{eq:FockDifferentialEntropyApprox}$ becomes a reasonable approximation only for $n \gg 1$.

\subsection{Mixture of number eigenstates}
Next, we analyze some particular mixtures of number eigenstates. For simplicity we concentrate on a mixed state of the first two number eigenstates, as well as a thermal state. We will again compare different entropic uncertainty relations and investigate how close they are to being violated.

\subsubsection{Mixture of first two number eigenstates}
\begin{figure*}[t!]
    \centering
    \includegraphics[height=0.22\textwidth]{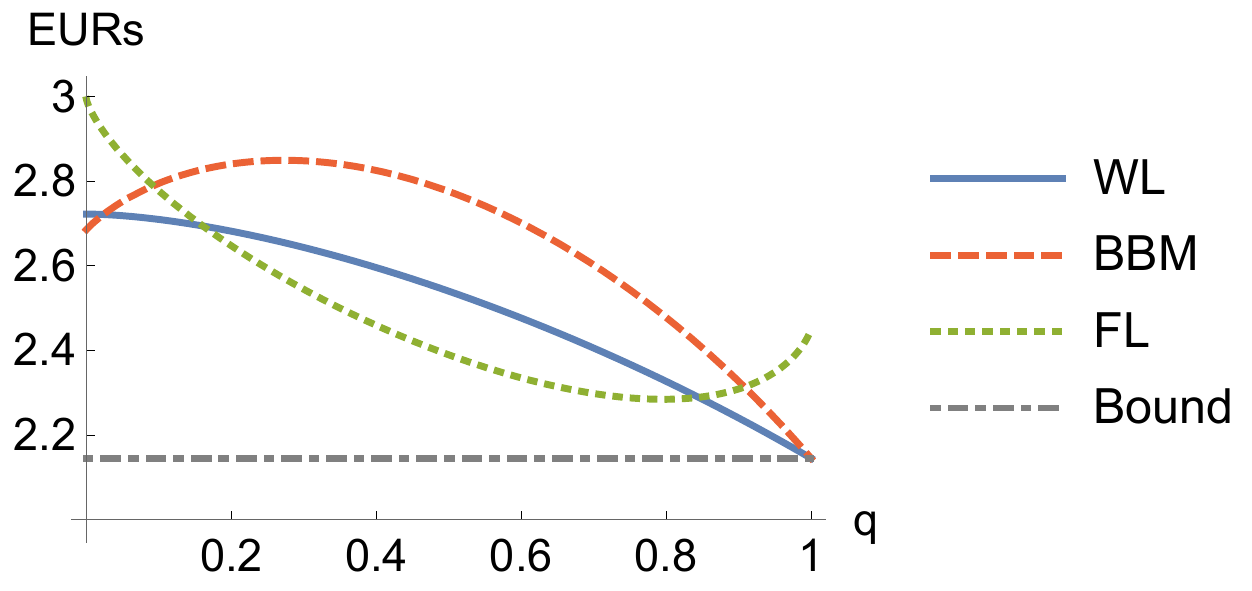}
    \includegraphics[height=0.22\textwidth]{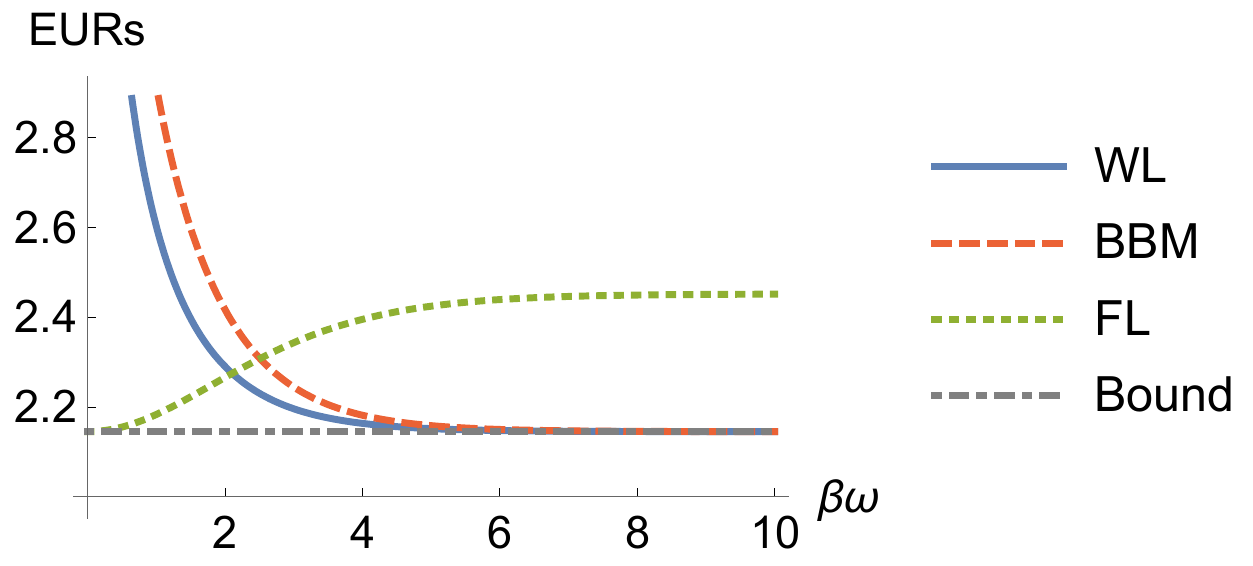}
    \caption{Left panel: Entropic uncertainty relations (EURs, cf. \eqref{eq:EURSummary}) for a mixture of the ground state with probability $q$ and the first excited state with probability $1-q$. The Wehrl entropy decreases monotonically for larger $q$, i.e. if the ground state contributes more to the total state. Right panel: Entropic uncertainty relations for the thermal state. The Wehrl-Lieb relations is again tighter than the Białynicki-Birula and Mycielski relation and both become tight for zero temperature $\beta \to \infty$. In contrast the Frank-Lieb (FL) relation becomes tight in the infinite temperature limit $\beta \to 0$, but is less tight for lower temperatures.} 
    \label{fig:QHOEURMixedStates}
\end{figure*}

A relatively simple mixed state is the following superposition of the first two number eigenstates,
\begin{equation}
    \rho_{0 1} = q \ket{0} \bra{0} + (1-q) \ket{1} \bra{1}.\label{eq:mixure01}
\end{equation}
The resulting entropies involve now a sum inside a logarithm. We use numerical methods to solve the integrals. The resulting curves for the entropic uncertainty relations are shown in \autoref{fig:QHOEURMixedStates} (left panel). Therein, one can see that the Wehrl-Lieb relation is tighter than the Białynicki-Birula and Mycielski relation except in a small region around $q=0$, where \eqref{eq:mixure01} corresponds to the first excited state. Thus we observe a similar situation as for pure states. Also, the Wehrl-Lieb relation turns out to be less concave than that of Białynicki-Birula and Mycielski. In contrast, the Frank-Lieb relation becomes tightest for more mixed states.

\subsubsection{Thermal state}
At last, we consider a thermal state $\rho_T$ under the assumption that the Hamiltonian is given by $H=\omega(n+1/2)$. The thermal state is then the state with maximal entropy for a fixed energy expectation value \cite{Jaynes1957,Jaynes19572,Jaynes1968,Landau1980}. It reads
\begin{equation}
    \rho_T = \frac{1}{Z} e^{- \beta H} = \sum_{n=0}^{\infty} p_n \ket{n} \bra{n},
\end{equation}
where $p_n$ is the Boltzmann distribution
\begin{equation}
    p_n = \frac{1}{Z} e^{- \beta \omega (n + \frac{1}{2})},
\end{equation}
and
\begin{equation}
    Z = \sum_{n=0}^{\infty} e^{- \beta \omega (n + \frac{1}{2})} = \frac{e^{- \beta \omega/2}}{1-e^{-\beta \omega}} = \frac{1}{2 \sinh(\beta \omega / 2)},
\end{equation}
is the canonical partition function. To calculate the probability density $f_T (x)$ we make use of Mehler's formula \cite{Mehler1866}, which leads to
\begin{equation}
    f_T (x) = \frac{1}{Z} \sum_{n=0}^{\infty} \abs{\psi_n (x)}^2 e^{- \beta \omega (n + \frac{1}{2})} = \frac{1}{\sqrt{2 \pi \sigma^2}} e^{- \frac{x^2}{2 \sigma ^2}},
\end{equation}
where we used the abbreviation
\begin{equation}
    \frac{1}{2 \sigma^2} = \tanh \left( \frac{\beta \omega}{2} \right).
\end{equation}
Thus, the distribution $f_T (x)$ is a Gaussian centered around $x=0$ with variance $\sigma^2$, which depends non-linearly on the inverse temperature $\beta$.
The associated differential entropy is
\begin{equation}
    h (f) = \ln \left(\sqrt{2 \pi e \sigma^2} \right).
\end{equation}
A similar consideration for the momentum distribution leads to the following expression for the uncertainty relation of Białynicki-Birula and Mycielski
\begin{equation}
    h (f_T) + h (g_T) = 1 + \ln \pi - \ln \left( \tanh \left( \frac{\beta \omega}{2} \right) \right).
\end{equation}
We see that for $\beta \to \infty$, i.\ e.\ in the zero temperature limit, the inequality becomes tight, as expected based on the discussion in section \ref{sec:NumberEigenstates}.

Since thermal states are highly mixed we should have a closer look at the Frank-Lieb relation. Additionally to the computation of the differential entropies, we need to calculate the von Neumann entropy of the thermal state $\rho_T$. It is given by
\begin{equation}
    S(\rho_T) = - \ln(1 - e^{- \beta \omega}) + \frac{\beta \omega}{e^{\beta \omega} - 1}.
\end{equation}
Thus, the left-hand side of the Frank-Lieb relation reads (again we rearrange everything such that the right-hand side equals $\ln e \pi$)
\begin{equation}
    \begin{split}
        &h (f_T) + h (g_T) - S(\rho_T) + 1 - \ln 2\\
        &= 2 + \ln \left( \frac{\pi}{2} \frac{1 - e^{- \beta \omega}}{\tanh( \beta \omega / 2)} \right) - \frac{\beta \omega}{e^{\beta \omega} - 1}.
    \end{split}
\end{equation}
Also, we need to calculate the Husimi $Q$-distribution, which yields
\begin{equation}
    \begin{split}
        Q_T (x,p) &= \frac{1}{Z} \sum_{n=0}^{\infty} Q_n (x,p) e^{- \beta E_n}\\
        &= \frac{1}{Z} e^{- \frac{1}{2} (x^2 + p^2) - \frac{\beta \omega}{2}} \sum_{n=0}^{\infty} \frac{1}{2^n n!} (x^2 + p^2)^n e^{- \beta  \omega n} \\
        &= \frac{1}{Z} e^{- \frac{1}{2} (x^2 + p^2) \left(1-e^{- \beta \omega} \right) - \frac{\beta \omega}{2}}.
    \end{split}
\end{equation}
For the left-hand side of the Wehrl-Lieb uncertainty relation we then find
\begin{equation}
    S_{\text{W}} (\rho_T) + \ln \pi = 1 + \frac{\beta \omega}{2} + \ln \left(\frac{\pi}{2} \text{csch} (\beta \omega/2) \right).
\end{equation}
We show a comparison of the three entropic uncertainty relations in \autoref{fig:QHOEURMixedStates} (right panel). The Wehrl-Lieb relation is again stronger than the Białynicki-Birula and Mycielski relation. This is due to the fact that for a thermal state the ground state is populated most and higher energy eigenstates are suppressed exponentially. Moreover, both become tight for the temperature tending to zero, which corresponds to the system approaching the ground state. Interestingly, the relation by Frank and Lieb behaves differently, which was already noted by the authors themselves \cite{Frank2012,Coles2017}. The uncertainty relation becomes tight in the \textit{infinite} temperature limit in contrast to the other two inequalities and it is more loose for low temperatures.

\subsubsection{General comment}
What we have observed is that the Frank-Lieb relation \eqref{eq:FrankEUR} is tightest for highly mixed states, while the Wehrl-Lieb relation \eqref{eq:WehrlLiebInequality} is tightest in the number eigenstate regime (except for $n=1$). Furthermore, we have seen that the Wehrl-Lieb relation is less concave under mixing than the Białynicki-Birula and Mycielski relation (cf. \autoref{fig:QHOEURMixedStates}). Thus, we can conjecture that the Wehrl-Lieb entropic uncertainty relation is tighter than the Białynicki-Birula and Mycielski entropic uncertainty relation in most situations. Furthermore, we have seen that the Wehrl entropy is very often easy to compute, especially compared to differential entropies of marginal distributions.

\section{Bipartite quantum system}
\label{sec:Bipartite}
In the second part of this work we consider a bipartite quantum system $AB$ with partitions $A$ and $B$ of $N$ and $M$ modes, respectively. Moreover, we associate the sets of coherent states $\ket{\vec \alpha}$ to $A$, $\ket{\vec \beta}$ to B and $\ket{\vec \alpha \vec \beta} = \ket{\vec \alpha} \otimes \ket{\vec \beta}$ to $AB$. Then, the local Husimi $Q$-distribution $Q^{\rho}_{A} (\vec \alpha)$ can be obtained from the global one $Q^\rho (\vec \alpha, \vec \beta)$ by integrating out system $B$
\begin{equation}
    Q^{\rho}_{A} (\vec \alpha) = \int \frac{\text{d}^{2M} \beta}{\pi^M} Q^{\rho} (\vec \alpha, \vec \beta).
\end{equation}
Note that this agrees with the Husimi $Q$-distribution obtained from the reduced density matrix $\rho_A = \text{Tr}_B\{\rho\}$. Similarly, one can define a local Husimi $Q$-distribution $Q^{\rho}_{B} (\vec \beta)$ for subsystem $B$.

It is also useful to consider a situation where a heterodyne measurement is performed first only on subsystem $B$. Assuming that the outcome is $\vec\beta$ leads to a Husimi $Q$-distribution for subsystem $A$ that we call the \textit{conditional} Husimi $Q$-distribution,
\begin{equation}
\begin{split}
    Q^{\rho}(\vec \alpha | \vec \beta) = & \Tr \left\{\frac{\braket{\vec \beta | \rho | \vec \beta}}{\braket{\vec \beta | \rho_B | \vec \beta}} E_{\vec \alpha} \right\} = \frac{\langle \vec \alpha \vec\beta | \rho | \vec\alpha \vec\beta \rangle }{\langle \vec \beta |\rho_B| \vec \beta \rangle} \\
    = & \frac{Q^{\rho}(\vec \alpha, \vec \beta) }{Q^{\rho}_{B} (\vec \beta)}.
\end{split}
\label{eq:ConditionalHusimiDef}
\end{equation}
Note that these relations are in complete analogy to classical probability densities.

\subsection{Wehrl relative entropy}
In the following we will consider different quantities used in classical and quantum information theory for Husimi $Q$-distributions defined in analogy to the Wehrl entropy $S_{\text{W}} (\rho)$. We begin with defining the Wehrl relative entropy, which will also be helpful to define a Wehrl conditional entropy and Wehrl mutual information. For any state $\rho$ and some model state $\sigma$ on an $N$-mode system it reads (in complete analogy to the standard definition by Kullback and Leibler \cite{Kullback1951,Kullback1959})
\begin{equation}
    \begin{split}
        S_{\text{W}} (\rho \| \sigma)
        = \int \frac{\text{d}^{2N} \alpha}{\pi^N} Q^\rho (\vec \alpha) (\ln Q^{\rho} (\vec \alpha) - \ln Q^\sigma (\vec \alpha)).\label{eq:generalWehrlRelativeEntropy}
    \end{split}
\end{equation}
The usual support condition for the states $\text{supp}(\rho) \subseteq \text{supp} (\sigma)$ then translates into the support condition for the Husimi $Q$-distributions $\text{supp}(Q^{\rho}) \subseteq \text{supp} (Q^{\sigma})$. If this condition is violated we set $S_{\text{W}} (\rho \| \sigma) = + \infty$. Intuitively, \eqref{eq:generalWehrlRelativeEntropy} can be interpreted as a measure for the \textit{distinguishability} of the true Husimi $Q$-distribution $Q^{\rho}$ and some model distribution $Q^{\sigma}$. Furthermore, it is a non-negative quantity being zero if and only if the two distributions agree\footnote{See appendix \ref{app:WehrlRelativeEntropy} for a simple proof.}. However, it should be emphasized that the most important property of the quantum relative entropy, namely its monotonicity under quantum channels, does \textit{not} hold for the Wehrl relative entropy. This is due to the fact the Wehrl entropy is not invariant under general unitary transformations. In fact, the Wehrl entropy is only invariant for transformations which map coherent states to coherent states \cite{Wehrl1979}.

\subsection{Wehrl conditional entropy}
Let us recall first some properties of quantum conditional entropies.
In the finite dimensional case there exist several ways of defining this quantity, which are all equivalent. For an infinite dimensional Hilbert space this is not necessarily the case. An adequate definition for both cases comprises the quantum relative entropy and requires $S(\rho_A) < \infty$ (see ref.\ \cite{Kuznetsova2011}). Then, we can define the quantum conditional entropy as
\begin{equation}
    S^\rho (A | B) = S (\rho_A) - S(\rho \| \rho_A \otimes \rho_B),
    \label{eq:QuantumConditionalEntropy}
\end{equation}
which reduces to
\begin{equation}
    S^\rho (A | B) = S (\rho) - S (\rho_B),
\end{equation}
provided that the von Neumann entropy of subsystem $B$ is also finite $S(\rho_B) < \infty$.

The quantum conditional entropy $S^\rho (A | B)$ allows for a simple separability criterion. In particular, one can easily prove that all separable states\footnote{Note that for an infinite dimensional Hilbert space a separable state may also be written as a convex integral over pure product states. This is particularly interesting for mixed Gaussian states.}
\begin{equation}
    \rho = \sum_i p_i \, \rho^i = \sum_i p_i \left(\rho_A^i \otimes \rho_B^i \right),
    \label{eq:SeparableState}
\end{equation}
where $p_i \ge 0$ and $\sum_i p_i = 1$ is a probability distribution and $\rho_A^i$ and $\rho_B^i$ can be taken to be pure, fulfill
\begin{equation}
    S^\rho (A | B) \ge 0.  
    \label{eq:QuantumConditionalEntropySeparable}
\end{equation}
Thus, the violation of the latter inequality is a sufficient condition for entanglement \cite{Cerf1997,Cerf1999}. One may easily construct examples where this bound is violated, e.\ g.\ a pure entangled state.

In complete analogy to \eqref{eq:QuantumConditionalEntropy} we define the Wehrl conditional entropy for $S_{\text{W}} (\rho_A) < \infty$ as
\begin{equation}
    S^\rho_{\text{W}} (A | B) = S_{\text{W}} (\rho_A) - S_{\text{W}} (\rho \| \rho_A \otimes \rho_B).
    \label{eq:WehrlConditionalEntropy}
\end{equation}
Analogously, we get
\begin{equation}
    S^\rho_{\text{W}} (A | B) = S_{\text{W}} (\rho) - S_{\text{W}} (\rho_B)
\end{equation}
if $S_{\text{W}} (\rho_B) < \infty$. 

Note that one may also write this with the conditional Husimi $Q$-distribution in \eqref{eq:ConditionalHusimiDef} as
\begin{equation}
\begin{split}
S^\rho_\text{W}(A | B) = & - \int \frac{d^{2M}\beta}{\pi^M} Q^\rho_{B}(\vec \beta) \\ & \times \int \frac{d^{2N}\alpha}{\pi^N} Q^\rho(\vec \alpha | \vec \beta ) \ln Q^\rho(\vec \alpha | \vec \beta),
\end{split}
\end{equation}
similar as for a classical conditional entropy. Because the conditional Husimi $Q$-distribution is itself a Husimi $Q$-distribution, the Wehrl-Lieb inequality implies 
\begin{equation}
    S^\rho_\text{W}(A | B) \geq N,
    \label{eq:ConditionalWL}
\end{equation}
which is a refined version of the monotonocity property \eqref{eq:WehrlMonotonicity} since it can be reformulated as
\begin{equation}
    S_{\text{W}} (\rho_B) + N \le S_{\text{W}} (\rho).
    \label{eq:WehrlRefinedMonotonicity}
\end{equation}
Also, the relation \eqref{eq:ConditionalWL} can be interpreted as an entropic uncertainty relation in the presence of classical memory (see \cite{dePalma2018c} for an entropic uncertainty relation with quantum memory). In contrast to the relation discussed in ref.\ \cite{dePalma2018c}, where the analysis was restricted to quasi-classical states and corresponding entropies, \eqref{eq:ConditionalWL} shows that a fully classical Wehrl conditional entropy can not serve as an entanglement witness analogous to eq.\ \eqref{eq:QuantumConditionalEntropySeparable}. In this sense, it behaves like a classical conditional entropy.

\subsection{Wehrl mutual information}
To quantify the correlations between the two subsystems $A$ and $B$ we introduce the Wehrl mutual information
\begin{equation}
    I^\rho_{\text{W}} (A : B) = S^\rho_{\text{W}} (\rho \| \rho_A \otimes \rho_B).
    \label{eq:WehrlMutualInformation}
\end{equation}
Most importantly, its non-negativity is inherited from the Wehrl relative entropy \eqref{eq:generalWehrlRelativeEntropy}. Also, it is zero if and only if the state $\rho$ is a product state. We can express \eqref{eq:WehrlMutualInformation} in terms of other Wehrl entropies provided that they are finite
\begin{equation}
    \begin{split}
        I^\rho_{\text{W}} (A : B) &= S_{\text{W}} (\rho_A) + S_{\text{W}} (\rho_B) - S_{\text{W}} (\rho) \\
        &=  S_{\text{W}} (\rho_A) - S^\rho_{\text{W}} (A | B).
    \end{split}
\end{equation}
In general, a mutual information can not distinguish between classical and quantum correlations. Thus, let us now consider the case where some state $\rho$ is known to be pure. In this case the Wehrl mutual information $I^\rho_{\text{W}} (A:B) $ quantifies only the quantum correlations between $A$ and $B$. Moreover, it is zero if and only if the state is a product state $\rho = \rho_A \otimes \rho_B$, which follows directly from the according property of the Wehrl relative entropy. Therefore, we can conclude that the Wehrl mutual information $I^\rho_{\text{W}} (A : B) $ is a perfect witness for pure state entanglement. 

It should be noted that the positive partial transpose (PPT) criterion already provides a necessary and sufficient condition for pure state entanglement (see e.g. \cite{Horodecki2009}), but requires the knowledge of the state $\rho$, which is typically inaccessible in experiments. Therefore, the advantage of the Wehrl mutual information $I^\rho_{\text{W}} (A : B) $ lies in its measurability and in this sense it can be regarded as a measurable equivalent to the PPT criterion for pure states. Also, one should note that not every classical mutual information is a perfect witness for pure state entanglement. For example, any marginal mutual information does not contain information about correlations in or with other variables.

Additionally, it is interesting to check whether $I^\rho_{\text{W}} (A:B) $ fulfills the requirements for an entanglement \textit{measure} or an entanglement \textit{monotone}. Since the Wehrl mutual information does not reduce to the entanglement entropy, it is not an entanglement measure. Also, the Wehrl mutual information does \textit{not} fulfill a minimum requirement, namely invariance under local unitaries
\begin{equation}
    I^{\rho^\prime}_{\text{W}} (A:B) \neq I^\rho_{\text{W}} (A : B),
\end{equation}
with $\rho^\prime = (U_A \otimes U_B) \, \rho\, (U_A^\dag \otimes U_B^\dag)$ and $U_{A,B}$ unitary. A simple counterexample is given by $U_A = \mathds{1}$ and $U_B = S(\kappa)$, where $S(\kappa)$ is the squeezing operator on subsystem $B$. Therefore, the Wehrl mutual information should be considered as a witness for pure state entanglement and not as an entanglement monotone.

Nevertheless, there is a non-trivial relation between the Wehrl mutual information and the quantum mutual information. Following \cite{Lieb2005}, we have the relation
\begin{equation}
    I^\rho_{\text{W}} (A : B) \le I^\rho (A : B).
    \label{eq:MutualInformationBound}
\end{equation}
For globally pure states the latter relation reduces to
\begin{equation}
    \frac{1}{2} I^\rho_{\text{W}} (A:B) \le S(\rho_A) = S(\rho_B),
\end{equation}
which is an interesting and useful relation when it comes to quantifying entanglement in an actual experiment. If the Husimi $Q$-distribution of a pure state $\rho$ is measured, one can not only decide if the state $\rho$ is entangled or not, one can also give a lower bound on the entanglement entropy $S(\rho_A)$. Thus, the measurable Wehrl mutual information $I^\rho_{\text{W}} (A : B)$ allows to make a statement about how much two subsystems are entangled \textit{at least}. If the global state $\rho$ is not pure, one can still make a statement about how much the two subsystems are correlated at least.

It should be noted that the left-hand side of the relation \eqref{eq:MutualInformationBound} is not restricted to the Wehrl mutual information, but holds true for all classical mutual informations constructed from positive operator-valued measures in subsystems $A$ and $B$ (cf. corollary 2 in ref.\ \cite{Lieb2005}).

Also, we want to point out that the relation \eqref{eq:MutualInformationBound} is similar to an open conjecture stated in ref.\ \cite{Schneeloch2014}, which is based on a homodyne detection protocol. It reads
\begin{equation}
    I^{\rho} (f_A : f_B) + I^{\rho} (g_A : g_B) \overset{?}{\le} I^{\rho} (A:B),
\end{equation}
where $f_A = f(x_A), f_B=f(x_B)$ are probability densities of the positions and $g_A = g(p_A)$, $g_B=g(p_B)$ are the momentum distributions measured in $A$ or $B$, respectively. Note that the measured mutual informations in the latter relation do not contain any information about correlations between positions and momenta, which are naturally included in the Wehrl mutual information $I^\rho_{\text{W}} (A : B) $.

\subsection{Entanglement of common pure states}
\label{sec:EntanglementExamples}
In the following, we compute the Wehrl mutual information $I^\rho_{\text{W}} (A : B)$ and the Wehrl conditional entropy $S^\rho_{\text{W}} (A | B)$ for some well-known pure states. To that end we use the parametrization \eqref{eq:PhaseToPositionMomentum} for each subsystem.

\subsubsection{Gaussian states}
We begin with a general Gaussian state $\rho$. Such a state is characterized by a global Husimi $Q$-distribution of the form\footnote{Without loss of generality we assume that the expectation values of the quadratures vanish.}
\begin{equation}
    Q^{\rho} (\vec{r}) = \frac{1}{Z} \exp \left(- \frac{1}{2} \, \vec{r}^T \cdot C \cdot \vec{r} \right),
    \label{eq:HusimiQGauss}
\end{equation}
where the bilinear form in the exponent contains all four phase space vectors $\vec x_A, \vec p_A, \vec x_B, \vec p_B$ combined into a single vector
\begin{equation}
    \vec{r} = \left(\vec{r}_A, \vec{r}_B \right)^T = \left(\vec x_A, \vec p_A, \vec x_B, \vec p_B \right)^T,
    \label{eq:rVector}
\end{equation}
and the symmetric matrix $C$ has the form
\begin{equation}
    C = 
    \begin{pmatrix}
    C_A & C_M \\
    C_M^T & C_B
    \end{pmatrix}.
    \label{eq:MMatrix}
\end{equation}
Note that $C$ is related to the covariance $V$ in symplectic phase space,
\begin{equation}
    V_{i j} = \frac{1}{2} \Tr \left\{\rho \left(r_i r_j + r_j r_i \right) \right\},
\end{equation}
by
\begin{equation}
    C = \left(V + \frac{1}{2} \, \mathds{1} \right)^{-1}.
\end{equation}
This follows from convolving the corresponding Gaussian Wigner $W$-distribution with a Gaussian distribution of covariance $(1/2) \mathds{1}$.

Also, $Z$ is a normalization constant, such that \eqref{eq:HusimiQNormalization} is fulfilled. Then, the expression \eqref{eq:HusimiQGauss} can be rewritten as
\begin{equation}
    \begin{split}
        &Q^{\rho} (\vec{r}_A, \vec{r}_B)\\
        &= \frac{1}{Z} \exp \left(- \frac{1}{2} \vec{r}_A^T C_A \vec{r}_A - \frac{1}{2} \vec{r}_B^T C_B \vec{r}_B -  \vec{r}_A^T C_M \vec{r}_B \right),
    \end{split}
\end{equation}
where the correlations are encoded in the mixing term $\vec{r}_A^T C_M \vec{r}_B$. If this term would be absent, the Husimi $Q$-distribution for the joint system $A B$ would factorize into the individual Husimi $Q$-distributions.

\begin{figure*}[t!]
    \centering
    \includegraphics[height=0.20\textwidth]{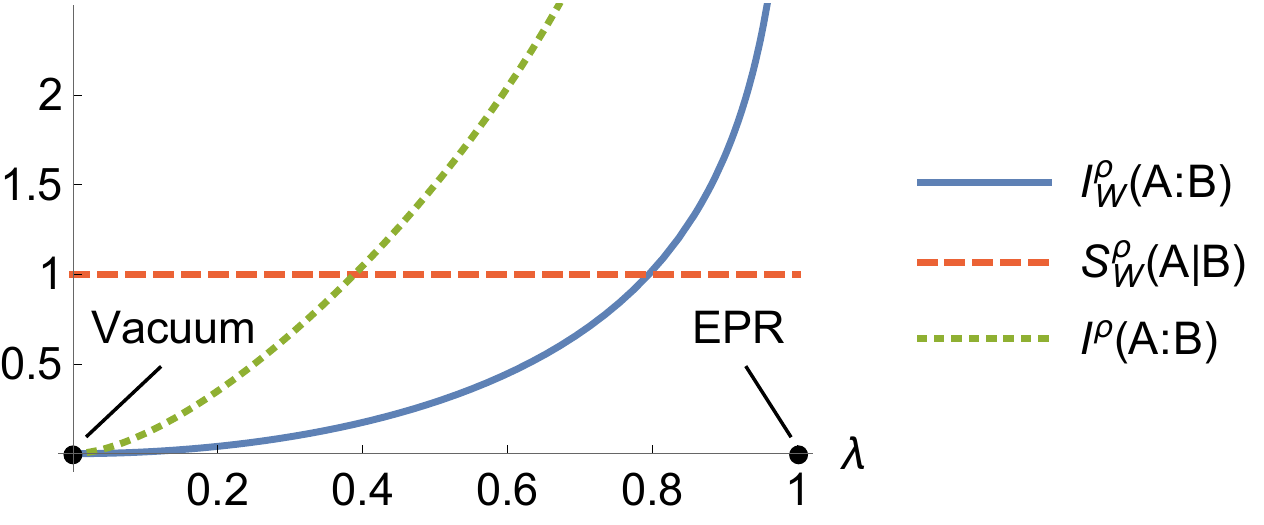}
    \includegraphics[height=0.20\textwidth]{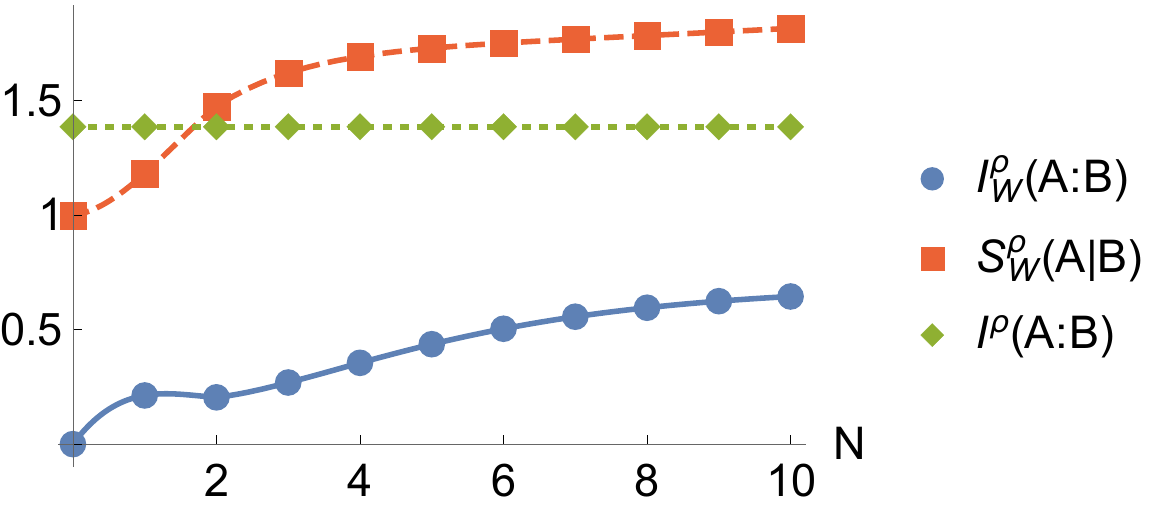}
    \caption{Left panel: Wehrl mutual information $I^\rho_{\text{W}} (A : B)$ and Wehrl conditional entropy $S^\rho_{\text{W}} (A | B)$ together with the quantum mutual information $I^\rho (A : B)$ of the two-mode squeezed state as a function of the squeezing parameter $\lambda$. For $\lambda = 0$ one recovers the product state of two vacua, while for $\lambda \to 1$ the state approaches the fully correlated Einstein-Podolsky-Rosen (EPR) state. Right panel: Wehrl mutual information $I^\rho_{\text{W}} (A : B)$, Wehrl conditional entropy $S^\rho_{\text{W}} (A | B)$ and quantum mutual information $I^\rho (A : B)$ for the N00N states up to $N=10$ (the curves represent interpolations). Higher N00N states have larger Wehrl mutual information, although the value for $N=1$ is slightly smaller than for $N=2$.} 
    \label{fig:QHOWehrlMutualInformation}
\end{figure*}

Performing a Gaussian integration, we arrive at the joint Wehrl entropy
\begin{equation}
    S_{\text{W}} (\rho) = - \frac{1}{2} \ln \det C + N + M.    
    \label{eq:WehrlEntropyGaussianAB}
\end{equation}
Note that the Wehrl-Lieb inequality for $N+M$ modes then implies that 
\begin{equation}
    \det C \le 1 \Leftrightarrow \det \left(V + \frac{1}{2} \, \mathds{1} \right) \ge 1, \label{eq:CovarianceUncertainty}
\end{equation}
which is an uncertainty relation in terms of the covariance matrix $V$ for the special case of Gaussian states.

Unlike the Robertson-Schrödinger uncertainty relation for Gaussian states $\det V \geq 1/2^{N+M}$ (see for example \cite{Hertz2019,Berges2018}), \eqref{eq:CovarianceUncertainty} is not invariant under similarity transformations of $V$ like symplectic transformations (e.g. squeezing). This shows a difference between the Robertson-Schrödinger uncertainty relation and the Wehrl-Lieb inequality. The former introduces a minimum for the uncertainty product while the latter is only tight for coherent states, i.e. equal uncertainty in position and momentum.

The next step is to compute the reduced Husimi $Q$-distribution for one subsystem. Without loss of generality we consider subsystem $B$ and end up with
\begin{equation}
    \begin{split}
        &Q^\rho_{B} (\vec{r}_B) \\
        &= \frac{1}{(2 \pi)^N} \int \text{d}^{2N} r_A \, Q^{\rho} (\vec{r}) \\
        &= \sqrt{\frac{\det C}{\det C_A}} \, \exp \left(- \frac{1}{2} \vec{r}_B^T \left(C_B - C_M^T C_A^{-1} C_M \right) \vec{r}_B \right).
    \end{split}
\end{equation}
Calculating the local Wehrl entropy gives
\begin{equation}
    S_{\text{W}} (\rho_B ) = - \frac{1}{2} \ln \det C + \frac{1}{2} \ln \det C_A + M,
    \label{eq:WehrlEntropyGaussianB}
\end{equation}
where we used the identity
\begin{equation}
    \det C = \det C_A \, \det \left(C_B - C_M^T C_A^{-1} C_M \right),
\end{equation}
which holds provided that $\det C_A \neq 0$. Here, the Wehrl-Lieb inequality \eqref{eq:WehrlLiebInequality} requires
\begin{equation}
    \det C \le \det C_A.
\end{equation}
Assuming that all Wehrl entropies are finite allows to calculate the two quantities of our interest. Both appear to have a simple form
\begin{equation}
    \begin{split}
        S^\rho_{\text{W}} (A | B) &= N - \frac{1}{2} \ln \det C_A, \\
        I^\rho_{\text{W}} (A : B) &= \frac{1}{2} \ln \frac{\det C_A \det C_B}{\det C}.
    \end{split}
    \label{eq:PureGaussianStateEntanglement}
\end{equation}
It can be seen easily that in the pure state case that $I^\rho_\text{W} (A : B) = 0$ is a necessary and sufficient separability condition for Gaussian states as it corresponds to the Wehrl relative entropy between the full Husimi $Q$-distribution and its factorization. The latter is equivalent to $C_M=0$, which itself is equivalent to $V_M = 0$. This already implies separability via the positive partial transpose (PPT) criterion for pure Gaussian states \cite{Lami2018,Serafini2017}. For the special case $N=M=1$, which is the setup of the Simon criterion \cite{Simon2000}, we show that the Simon criterion indeed implies $C_M = 0$ for pure separable Gaussian states in appendix \ref{app:SimonCriterion}. At least in this special case, the criterion of ref. \cite{Simon2000} and our condition for separability, $I^\rho_\text{W} (A : B) = 0$, are therefore equivalent.

Interestingly, we find also that the Wehrl conditional entropy $S^\rho_{\text{W}} (A | B)$ is independent of the correlations encoded in $C_M$. Furthermore, applying \eqref{eq:ConditionalWL} to \eqref{eq:PureGaussianStateEntanglement} leads to local uncertainty relations of the form
\begin{equation}
    \det C_A \le 1 \hspace{0.5cm} \text{and} \hspace{0.5cm} \det C_B \le 1.
\end{equation}

\subsubsection{Two-mode squeezed state}
As a more concrete example, we consider the two-mode squeezed state for $N=M=1$, which is defined by \cite{Weedbrook2012}
\begin{equation}
    \ket{\psi} = \sqrt{1 - \lambda^2} \sum_{n=0}^{\infty} (- \lambda)^n \ket{n_A,n_B},
    \label{eq:TwoModeSqueezedState}
\end{equation}
where the squeezing parameter $\lambda \in [0,1]$ is related to the squeezing parameter $r \in [0, \infty)$ by
\begin{equation}
    \lambda = \tanh r.
\end{equation}
Since the two-mode squeezed state is a pure Gaussian state, the corresponding Husimi $Q$-distribution is of Gaussian form \eqref{eq:HusimiQGauss}. Determining the matrix $C$ yields
\begin{equation}
    C_A = C_B = \mathds{1}_2 \hspace{0.5cm} \text{and} \hspace{0.5cm} C_M = \text{diag} (\lambda, - \lambda).
\end{equation}
As expected, the correlation matrix $C_M$ becomes zero for $\lambda = 0$, in which case the state \eqref{eq:TwoModeSqueezedState} is separable, such that the global Husimi $Q$-distribution factorizes. For the Wehrl conditional entropy and the Wehrl mutual information (cf. Eq. \eqref{eq:PureGaussianStateEntanglement}) we find
\begin{equation}
    \begin{split}
        S^\rho_{\text{W}} (A | B) &= 1, \\
        I^\rho_{\text{W}} (A : B) &= - \ln \left(1 - \lambda^2 \right).
    \end{split}
    \label{eq:TwoModeSqueezedStateEntanglement}
\end{equation}
Both quantities allow for a straight-forward classification in an actual experiment: since the Wehrl conditional entropy is independent of quantum correlations encoded in the squeezing parameter $\lambda$, any measured deviation from $1$ is sufficient to show that the state exhibits additional classical correlations. For example, the squeezing parameter $\lambda$ may fluctuate in subsequent realizations of an experiment, which increases the Wehrl conditional entropy due to its concavity. Note that this statement is a consequence of the Wehrl conditional entropy \eqref{eq:PureGaussianStateEntanglement} being independent of the correlations encoded in $C_M$. Also, it is interesting to note that the lower bound \eqref{eq:ConditionalWL} \textit{is} attained for the two-mode squeezed state (see \autoref{fig:QHOWehrlMutualInformation}, left panel, red dotted curve), to indicate the amount of quantum correlations captured by the Wehrl mutual information.

Furthermore, the Wehrl mutual information allows to set a lower bound on the amount of bipartite entanglement between $A$ and $B$. The latter is monotonically increasing for an increasing squeezing parameter (see \autoref{fig:QHOWehrlMutualInformation}, left panel, blue straight curve). In particular, it vanishes if and only if the state is separable ($\lambda \to 0$) and tends to infinity if the state approaches the Einstein-Podolsky-Rosen state ($\lambda \to 1$) representing perfect correlations $x_A = x_B$ and $p_A = - p_B$. Furthermore, we see how the bound \eqref{eq:MutualInformationBound} behaves as we have also plotted the quantum mutual information (see \autoref{fig:QHOWehrlMutualInformation}, left panel, green dotted curve), which can be calculated easily following refs.\ \cite{Serafini2003,Weedbrook2012}.

\subsubsection{N00N states}
Another interesting class of pure entangled states are the N00N states, which are of the form (the normalization constant is chosen such that the following expression is normalized to unity also for $N=0$)
\begin{equation}
    \ket{\psi_N} = \frac{1}{\sqrt{2 (1 + \delta_{0 N})}} \left(\ket{N,0} + \ket{0,N} \right),
\end{equation}
for $1+1$ modes (note that here $N$ denotes the $N$th excitation instead of the $N$th mode). These states are separable if and only if $N=0$, in which case we end up with two vacua again. Furthermore, for $N=1$ one recovers one of the four Bell states as a special case. 

The N00N states are known to be hard to detect if considered in the context of a continuous variables quantum system \cite{Weedbrook2012}. For example, the strong entropic separability criterion in ref.\ \cite{Walborn2009} only detects N00N states up to $N=5$. Since the Wehrl mutual information is a perfect witness, all N00N states are witnessed.

Calculating the global and one local Husimi $Q$-distribution gives
\begin{equation}
    \begin{split}
        Q^{\rho} (\vec{r}) = & \frac{1}{2^{N+1} \, N! \, (1 + \delta_{0 N})} \, e^{-(r_A^2 + r_B^2) /2} \\
        & \times \left( (x_A - i p_A)^N + (x_B - i p_B)^N \right) \\ 
        & \times \left( (x_A + i p_A)^N + (x_B + i p_B)^N \right), \\
        Q^\rho_{B} (\vec{r_B}) = & \frac{1}{2^{N+1} \, N!} \, e^{-r_B^2 /2} \, \left( r_B^{2 N} + 2^N N! \right),
    \end{split}
\end{equation}
where we used the abbreviations $r_i^2 = \abs{\vec{r}_i}^2 = x_i^2 + p_i^2$ for $i \in \{ A, B \}$. The Wehrl conditional entropy and the Wehrl mutual information can be computed numerically (see \autoref{fig:QHOWehrlMutualInformation}, right panel, same color scheme as before). We observe that the Wehrl mutual information increases monotonically for $N \ge 2$. It can be seen that it approaches the quantum mutual information, which is simply $I^{\rho} (A:B) = 2 \ln 2$ for all $N \in \mathbb{N}$.

\section{Conclusion and outlook}
\label{sec:Conclusion}
In summary, we have investigated here the relation between the Wehrl-Lieb entropic uncertainty relation for the Husimi $Q$-distribution and other relations for marginal distributions. Moreover, we have discussed to which extent the Wehrl entropy and related quantities are able to witness entanglement of continuous variables.

We found that the Wehrl-Lieb inequality is tighter than the Białynicki-Birula and Mycielski relation almost everywhere. In particular, we have shown its superior tightness for number eigenstates $\rho_n = \ket{n} \bra{n}$, where $n=1$ turned out to be an exception. Also, the calculation showed that Wehrl entropies are technically simpler to handle compared to differential entropies of marginal distributions.

In the second half, we have introduced analogs of well-known quantum information theoretic quantities in terms of Wehrl entropies. It turned out that especially the Wehrl mutual information is of great use when it comes to entanglement witnessing. Although it is neither an entanglement measure, nor an entanglement monotone, it is a \textit{perfect} witness for pure state entanglement, i.e. it detects \textit{all} pure entangled states. Furthermore, it gives a lower bound on the entanglement entropy, which allows to quantify to which extent some pure state $\rho = \ket{\psi} \bra{\psi}$ is entangled \textit{at least}.

To exemplify the potential of the Wehrl mutual information we discussed several examples. For the Gaussian state case, after a straight forward exercise in Gaussian integration, we found a simple formula allowing for a direct classification. Here it would be of further interest to relate the Wehrl mutual information as perfect witness for pure Gaussian states to existing criteria like for example the Simon criterion \cite{Simon2000}. Additionally, we have seen that the Wehrl-Lieb inequality for $N+M$ mode Gaussian states implies a novel uncertainty relation in terms of the covariance matrix, which may be studied further in the future. Another interesting example we considered was the class of $N00N$ states, which are particularly hard to detect, especially for large $N$. Also in this case the Wehrl mutual information was a reasonable entanglement witness, in fact it is a witness for all $N \in \mathbb{N}$.

Besides the Wehrl mutual information we also considered the Wehrl conditional entropy, which turned out to be bounded from below similar to the Wehrl entropy itself. This statement can be interpreted as an entropic uncertainty relation in the presence of classical memory as well as a refined version of the monotonicity property of the Wehrl entropy under partial trace. Furthermore, for the special case of Gaussian states, we found that the Wehrl conditional entropy is independent of the correlations encoded in the off-diagonal blocks of the covariance matrix.

Overall, we want to emphasize again that the significance of the Wehrl entropy and all related quantities is that they are \textit{measurable} when making use of the heterodyne detection scheme. Thus, the Wehrl mutual information and the Wehrl conditional entropy can be applied immediately to actual experiments, where they serve as a perfect witness for pure state entanglement and as an indicator for non-Gaussianity, respectively.

One question left over for future work concerns a more general notion of coherent states, namely $SU(N)$ coherent states. It should be checked whether the presented approach to witness pure state entanglement can be generalized to $SU(N)$ coherent states. This question is of particular interest for e.g. experiments with spinor Bose-Einstein condensates \cite{Kunkel2018,Bergh2021a,Bergh2021b}. 

It is noteworthy, that the Wehrl-Lieb inequality holds also for generalized coherent states (see refs.\ \cite{Davies1976,Grabowski1984}), which may be another direction for future investigations.

Other directions for subsequent research are the generalization of the presented quantities to quantify entanglement also in the multipartite case and to other interesting states, e.\ g.\ Schrödinger cat states (see \cite{Miranowicz2001} for results on the Wehrl entropy in this case).

Furthermore, it would be of great interest to generalize the concept of an entropy in quantum mechanical phase space to a quantum field theory. In this way, it should be possible to formulate entropic uncertainty relations for conjugate quantum fields with a Wehrl (relative) entropy and to constrain measurable (quantum) correlations analogous to what we have discussed here.

\section*{Acknowledgements}
The authors thank Martin Gaerttner and Markus Oberthaler for valuable discussions. This work is supported by the Deutsche Forschungsgemeinschaft (DFG, German Research Foundation) under Germany's Excellence Strategy EXC 2181/1 - 390900948 (the Heidelberg STRUCTURES Excellence Cluster), SFB 1225 (ISOQUANT) as well as FL 736/3-1.

\appendix

\section{Non-negativity of Wehrl relative entropy}
\label{app:WehrlRelativeEntropy}
Although the Husimi $Q$-distribution is only a \textit{quasi}-probability distribution, it is non-negative and normalized, which suffices to apply Jensen's inequality \cite{Wilde2013}. Because $- \ln x$ is a convex function for real positive $x$ we get
\begin{equation}
    \begin{split}
        S_{\text{W}} (\rho \| \sigma) &= - \int \frac{\text{d}^{2N} \alpha}{\pi^N} Q^{\rho} (\vec \alpha) \ln \frac{Q^{\sigma} (\vec \alpha)}{Q^{\rho} (\vec \alpha)} \\
        &\ge - \ln \left( \int \frac{\text{d}^{2N} \alpha}{\pi^N} Q^{\rho} (\vec \alpha) \frac{Q^{\sigma} (\vec \alpha)}{Q^{\rho} (\vec \alpha)} \right) \\
        &= - \ln \left( \int \frac{\text{d}^{2N} \alpha}{\pi^N} Q^{\sigma} (\vec \alpha) \right) = 0,
    \end{split}
\end{equation}
where the validity of the support condition was assumed in the third step.

Moreover, equality holds iff the two Husimi $Q$-distributions agree since $- \ln x$ is \textit{strictly} convex for $x>0$.

\section{Concavity of Wehrl conditional entropy}
\label{app:WehrlConditionalEntropy}
Writing out the Wehrl conditional entropy
\begin{equation}
    \begin{split}
        &S^\rho_{\text{W}} (A | B) \\
        &= - \int \frac{\text{d}^{2N} \alpha \, \text{d}^{2M} \beta}{\pi^{N+M}} Q^{\rho} (\vec \alpha, \vec \beta) \left(\ln Q^{\rho} (\vec \alpha, \vec \beta) - \ln Q^{\rho}_B (\vec \beta) \right)
    \end{split}
\end{equation}
reveals that the integrand is given by the simple function $f(x,y) = - x \ln x + x \ln y$ for real positive $x$ and $y$. This function is jointly concave in $x$ and $y$, which can easily be proven using Lieb's concavity theorem \cite{Nielsen2010}. Then, the monotonicity of the integral implies concavity for the Wehrl conditional entropy.

\section{Wehrl mutual information and Simon criterion}
\label{app:SimonCriterion}
We wish to show that the Simon criterion implies in the notation of eq. \eqref{eq:MMatrix} $C_M = 0$ for pure separable Gaussian states. To that end, let $V$ be the covariance matrix of a $(1+1)$-mode pure Gaussian state, such that we can write $V= (1/2) S^T S$ with $\det V = 1/16$ and where $S$ is a symplectic matrix $S \in Sp (4,\mathbb{R})$. $V$ can be represented as $S'^T V_0 S'$ with $S' = \text{diag} (S_1, S_2)$ being local symplectic transformations and $V_0$ denoting the normal form of $V$, 
\begin{equation}
    V_0 = \begin{pmatrix}
    a & 0 & c_1 & 0 \\
    0 & a & 0 & c_2 \\
    c_1 & 0 & b & 0 \\
    0 & c_2 & 0 & b
    \end{pmatrix}.
\end{equation}
Using that symplectic transformation have unit determinant, $\det S = \det S^T = 1$, the purity condition can be re-expressed as
\begin{equation}
    (a b - c_1^2) (a b - c_2^2) = \frac{1}{16}.    
\end{equation}
The parameters $a, b, c_1$ and $c_2$ can be further constrained from the symplectic eigenvalues $\lambda_i$ of $V_0$, which have to fulfill $\lambda_i = 1/2$ for pure Gaussian states. The symplectic eigenvalues also appear in pairs $(\pm \lambda_i)$ when considering the eigenvalues of $i \Omega V_0$, where $\Omega = \mathds{1} \otimes J$ with 
\begin{equation}
    J = \begin{pmatrix}
    0 & 1 \\
    -1 & 0
    \end{pmatrix},
\end{equation}
is the symplectic form, leading to another purity condition
\begin{equation}
    a^2 + b^2 + 2 c_1 c_2 = \det V_A + \det V_B + 2 \det V_M = \frac{1}{2}.
\end{equation}
Note that the latter conditions contains the invariants $I_1 = \det V_A, I_2 = \det V_B, I_3 = \det V_M$ and $I_4 = \Tr (J V_A J V_M J V_B J V_M^T)$ under the local symplectic transformations $S'$ and that $\det V = I_1 I_2 + I_3^2 - I_4$ \cite{Simon2000}.

Applying the PPT criterion, which corresponds to a mirror reflection in one of the two subsystems, to the covariance matrix $V_0$ of a pure Gaussian state leads to a new covariance matrix $\tilde{V}_0$. Assuming that the PPT condition holds, $\tilde{V}_0$ also needs to fulfill the latter purity condition. We find
\begin{equation}
    a^2 + b^2 - 2 c_1 c_2 = \det V_A + \det V_B - 2 \det V_M = \frac{1}{2},  
\end{equation}
as the mirror reflection only flips the sign of $\det \tilde{V}_M = - \det V_M$ and thus $\det V = \det \tilde{V}$. Therefore, the PPT criterion implies $\det V_M = c_1 c_2 = 0$. Without loss of generality we proceed with assuming that $c_2 = 0$. Then, the two purity conditions together with the non-negativity of $a b$ imply
\begin{equation}
    f(a) = a^2 \left( \frac{1}{2} - a^2 \right) \ge 16.
\end{equation}
Since $f(a)$ attains a global maximum for $a = 1/2$ with $f(a=1/2) = 1/16$, we can conclude that $a = b = 1/2$ and therefore $c_1 = 0$, implying $V_M = C_M = 0$ for every pure separable Gaussian state.

%%%%%%%% Bibliography %%%%%%%%

\vspace{1cm}

\bibliography{references.bib}

\end{document}